\documentclass[9pt,shortpaper,twoside,web]{ieeecolor2}
\usepackage{generic}
\usepackage{cite}
\usepackage{amsmath,amssymb,amsfonts}
\usepackage{algorithmic}
\usepackage{graphicx}
\usepackage{textcomp}
\usepackage{amsfonts}

\usepackage{amsmath}
\usepackage{booktabs}
\usepackage{subfig}
\usepackage{multirow}
\usepackage{tabularx}
\usepackage{array}
\usepackage{xcolor}

\def\BibTeX{{\rm B\kern-.05em{\sc i\kern-.025em b}\kern-.08em
    T\kern-.1667em\lower.7ex\hbox{E}\kern-.125emX}}
\begin{document}
\title{Effects of Muscle Synergy during Overhead Work with a Passive Shoulder Exoskeleton: A Case Study}
\author{Jin Tian, \IEEEmembership{Graduate Student Member, IEEE}, Baichun Wei, Chifu Yang, Suo Luo, Jiadong Feng, Ping Li, Changbing Chen, Yingjie Liu, Haiqi Zhu, and Chunzhi Yi, \IEEEmembership{Member, IEEE}
\thanks{This work is founded in part by National Natural Science Foundation of China (No. 62306083), the Postdoctoral Science Foundation of Heilongjiang Province of China (LBH-Z22175) and the Innovation Fund Project of National Commercial Aircraft Manufacturing Engineering Technology Research Center (COMAC-SFGS-2024-697). \textit{(Jin Tian and Baichun Wei contributed equally to this work.) (Correspondence authors: Haiqi Zhu, Chunzhi Yi)}}
\thanks{Jin Tian and Chifu Yang are with School of Mechatronics Engineering, Harbin Institute of Technology, Harbin, 150001, China (e-mail:	jin.tian@stu.hit.edu.cn; cfyang@hit.edu.cn).}
\thanks{Baichun Wei, Haiqi Zhu, and Chunzhi Yi are with the School of Medicine and Health, Harbin Institute of Technology, Harbin, 150001, China (e-mail: bcwei@hit.edu.cn; haiqizhu@hit.edu.cn; chunzhiyi@hit.edu.cn).}
\thanks{Suo Luo, Jiadong Feng and Ping Li are with the  Commercial Aircraft Corporation of China Ltd, Shanghai, 201323, China (e-mail: luosuo999999999@163.com; zjsxlzfjd123@163.com; 3220532757@qq.com).}
\thanks{Changbing Chen and Yingjie Liu are with the Emergency Science Research Institute, Chinese Institute of Coal Science, Beijing, 100013, China (e-mail: stark\_ccb@163.com; liuyingjie@mail.ccri.ccteg.cn).}}
\maketitle

\begin{abstract}
\textcolor{cyan}{\textit{Objective:}} Shoulder exoskeletons can effectively assist with overhead work. However, their impacts on muscle synergy remain unclear. The objective is to systematically investigate the effects of the shoulder exoskeleton on muscle synergies during overhead work. \textcolor{cyan}{\textit{Methods:}} Eight male participants were recruited to perform a screwing task both with (Intervention) and without (Normal) the exoskeleton. Eight muscles were monitored and muscle synergies were extracted using non-negative matrix factorization and electromyographic topographic maps. \textcolor{cyan}{\textit{Results:}} The number of synergies extracted was the same (n = 2) in both conditions. Specifically, the first synergies in both conditions were identical, with the highest weight of AD and MD; while the second synergies were different between conditions, with highest weight of PM and MD, respectively. As for the first synergy in the Intervention condition, the activation profile significantly decreased, and the average recruitment level and activation duration were significantly lower (p\textless0.05). The regression analysis for the muscle synergies across conditions shows the changes of muscle synergies did not influence the sparseness of muscle synergies (p=0.7341). In the topographic maps, the mean value exhibited a significant decrease (p\textless0.001) and the entropy significantly increased (p\textless0.01). \textcolor{cyan}{\textit{Conclusion:}} The exoskeleton does not alter the number of synergies and existing major synergies but may induce new synergies. It can also significantly decrease neural activation and may influence the heterogeneity of the distribution of monitored muscle activations. \textcolor{cyan}{\textit{Significance:}} This study provides insights into the potential mechanisms of exoskeleton-assisted overhead work and guidance on improving the performance of exoskeletons.
\end{abstract}

\begin{IEEEkeywords}
Shoulder exoskeleton, electromyography, overhead work, muscle synergy, non-negative matrix factorization.
\end{IEEEkeywords}

\section{Introduction}
\label{sec:introduction}
In the face of worsening global labor shortages and aging issues, the health of workers is increasingly crucial for individuals, companies, and society \cite{cieza2020global} \cite{enoka2008muscle}. Shoulder work-related musculoskeletal disorders (WMSDs) are one of the primary threats, often occurring when individuals are exposed to prolonged high physical loads or uncomfortable postures \cite{govaerts2021prevalence}. Overhead work, commonly involved in the industrial sector, has been identified as a major factor in developing shoulder WMSDs \cite{grazi2020design}. Despite the acceleration of the automation process and increased use of robots in factories, robots still cannot replace humans completely, especially in overhead tasks that require cognition and collaboration \cite{huysamen2018evaluation}. To reduce the physical burden of the human shoulder, shoulder exoskeleton offers a potential solution \cite{bar2021influence} \cite{howard2020industrial}. In particular, passive shoulder exoskeletons essentially rely on storing elastic potential energy in springs to compensate for shoulder gravity \cite{reyes2023shoulder}. Thus such exoskeletons would benefit from their characteristics of low cost, lightweight, and compactness compared with active shoulder exoskeletons, and have been accepted by several companies, such as BMW and Ford \cite{kim2022passive} \cite{de2023passive}.

Recent studies have shown that wearing shoulder exoskeletons can effectively reduce shoulder load, thus reducing the risk of shoulder WMSDs \cite{ojelade2023three} \cite{van2022exo4work} \cite{jorgensen2022influence} \cite{de2020passive} \cite{iranzo2020ergonomics}. Previous studies revealed that shoulder exoskeleton can reduce physical load on skeleton during arm elevation \cite{kim2018assessing} \cite{fritzsche2021assessing}. De Bock et al. for the first time demonstrated that a shoulder exoskeleton can reduce muscle activity and fatigue during overhead work \cite{de2022occupational}. Maurice et al. investigated the effects of a shoulder exoskeleton on both objective (e.g., muscle activation, movement duration, movement of the tool and arm) and subjective (e.g., acceptance) aspects during overhead work in a lab setting. Their results showed that their shoulder exoskeleton can delay the occurrence of shoulder WMSDs without affecting production efficiency \cite{maurice2019objective}. Pacifico et al. evaluated the impact of the shoulder exoskeleton on experienced workers in both on-site and simulated environments using electromyographic (EMG) activity and perceived exertion \cite{pacifico2022exoskeletons}. They demonstrated that their exoskeleton can reduce shoulder muscle activity with high usability and acceptance \cite{pacifico2022exoskeletons}. Although existing studies have presented the benefits of the shoulder exoskeleton, the evaluation method focuses on individual muscles rather than the coordination of multiple muscles, especially considering the fact that lower-limb exoskeletons have already shown their impact on muscle coordination \cite{li2019muscle} \cite{zhu2021effects} \cite{jeong2023muscle} \cite{liang2023surface}. 

\begin{figure*}[!t]
	{\includegraphics[width=7.7cm]{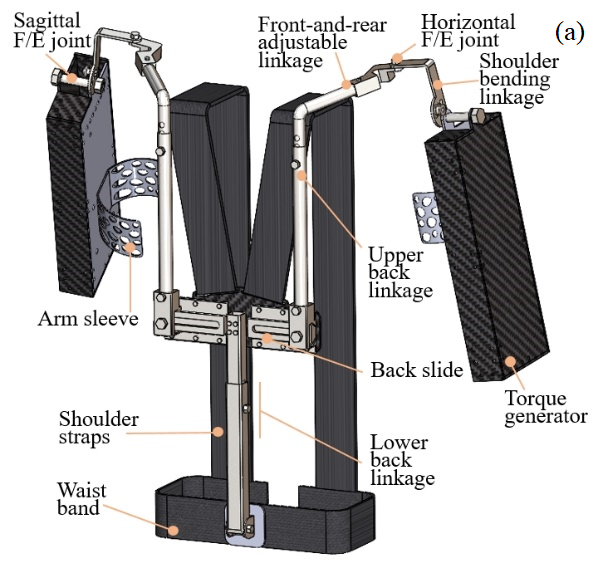}\label{fig1a}}
	{\includegraphics[width=11cm]{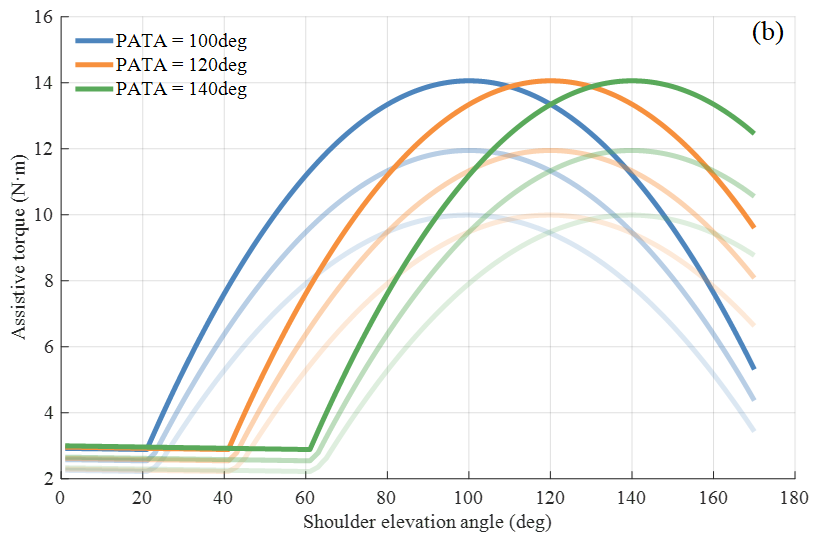}\label{fig1b}}
	\caption{The overview of the HIT-POSE. (a) The structure diagram of the exoskeleton. (b) The assistive torque profiles with three different PATAs and levels of assistance. Different colored profiles have different PATAs (i.e., 100deg, 120deg, and 140deg, respectively), and profiles of the same color but different transparency have different levels of assistance (i.e., 10N·m, 12N·m, and 14N·m, respectively).}
	\label{fig1}
\end{figure*}

The coordination of multiple muscles, i.e. muscle synergy, are thought to reflect motor control and the neuromuscular adaptation of the motor system \cite{safavynia2011muscle} \cite{wang2021synergy}. During exoskeleton-assisted overhead tasks, muscle synergy analysis can complement analyzing EMG of individual muscles, revealing whether the neural commands required for tasks have changed, or whether the contribution of muscles in each synergy has altered. Previous studies utilizing muscle synergy-based analysis have shown that lower-limb exoskeletons influenced the motor control of multiple muscles and can change the coordinate pattern of the muscles and the neural drive that activates the muscles \cite{jacobs2018motor} \cite{escalona2020effects} \cite{tan2018lateral}. Afzal et al. reported muscle synergies during exoskeleton-assisted walking in persons with multiple sclerosis \cite{afzal2022evaluation}. The results showed that the exoskeleton did not alter the existing muscle synergies, but can induce a new synergy. Li et al. performed muscle synergy analysis during exoskeleton-assisted sit-to-stand movements in subacute stroke survivors \cite{li2023exoskeleton}, and pointed out that the exoskeleton can shift the abnormal synergies of the paretic leg towards the normal synergies and increase the synergies of non-paretic leg. As the shoulder is a complex joint, and overhead tasks involve the coordinated activation of multiple muscles around the shoulder \cite{renda2023effects}, it is essential to investigate muscle synergy during overhead work. Our study aims to fill in the research gap by analyzing muscle synergies during exoskeleton-assisted overhead work, and thus investigating how the shoulder exoskeleton for assisting overhead work impacts neuromuscular coordination.

In terms of muscle synergy, non-negative matrix factorization (NMF) \cite{ghislieri2020muscle} and EMG topographic map \cite{jiang2021mapping} are two primary analyzing methods. NMF factorizes the activation of multiple muscles into the weight and timing coefficient matrices, denoting to what degree each muscle contributes to a synergy and the temporal activation of the synergies, respectively. Previous study \cite{rabbi2020non} has demonstrated that NMF performs the best among the commonly used factorization methods, such as principal component analysis, independent component analysis, and factor analysis. Therefore, NMF is widely used for muscle synergy analysis in exoskeleton-assisted movements \cite{tan2019muscle} \cite{tan2020differences}. However, this method of synergy analysis could be affected by physical constraints \cite{steele2015consequences} and experimental conditions \cite{jiang2024effect}, which could lead to an inaccurate and unstable conclusion. The EMG topographic map, the other tool for visualizing muscle synergy \cite{wang2021synergy}, can present a continuous visualization of muscle synergy and can complement the muscle synergy analysis based on NMF from a topographic perspective \cite{hu2010lumbar}. At the same time, the ability of the EMG topographic map to evaluate the effect of exoskeletons has been shown in a previous study \cite{jiang2024effect}. To obtain a more comprehensive and reliable temporal and spatial distribution of EMG as well as the influence of the exoskeleton on motor patterns during overhead work, we employed both NMF and EMG topographic map in our analysis.

The objective of this study was to systematically how shoulder exoskeleton for assisting overhead work can influence the muscle synergies, i.e. the spatial weight of muscles within each synergy and the temporal order of the activation of each synergy. Specifically, we used a passive shoulder exoskeleton named HIT-POSE to assist overhead work, which has already been designed and validated its kinematic compliance and assistance effectiveness \cite{tian2024novel}. Then, we used both NMF and EMG topographic map to investigate whether there were changes in the number and structure of synergies, merging and fraction of synergies and spatiotemporal shift of multi-channel muscle activation compared between with and without exoskeleton assistance.

\section{Methods}
To explore the potential impact of exoskeleton-assisted overhead work on muscle synergies, we formulated this section including the device involved, experimental setting, data analysis methodologies, and evaluation indicators.

\subsection{HIT-POSE: A Novel Passive Occupational Shoulder Exoskeleton}

The device under evaluation is the HIT-POSE, a novel passive occupational shoulder exoskeleton (Fig. 1(a)). The HIT-POSE provides assistive torque by compensating for the partial gravity of the shoulder. The device, connected to the user through the interface (i.e., shoulder straps, arm sleeves, and waist belt), weighs 2.8kg and has two degrees of freedom unilaterally. The shoulder structure comprises of the shoulder bending linkage, the front-and-rear adjustable linkage, and horizontal and sagittal joints. In addition, the torque generator is equipped with tension springs and a peak assistive torque angle (PATA) adjustment module to provide efficient assistance. It can adjust the assistive torque profile at both peak assistive torque and PATA levels to adapt to different tasks and individuals (Fig. 1(b)). Because various individuals and tasks may have different target task angles \cite{tian2024novel}, the exoskeleton is designed with the capability of increasing assistive torque when lifting the arm and decreasing the assistive torque when lowering the arm. The exoskeleton can adjust its size to accommodate different individuals (Fig. 1(a)). In the experiment, the PATA of the exoskeleton was adjusted to match the target task angle for each participant. The assistance is determined in a subject-specific way. The magnitude of the assistance is tuned according to user experience and feedback through a familiar phase of using exoskeleton before data collection.

\subsection{Participants and Experimental Protocol}

Eight healthy participants (all male, all right-handed, age: 25.4 ± 1.7 years, height: 176 ± 4.6 cm, weight: 68.7 ± 8.2 kg) were recruited for this study. All participants had no prior experience with the exoskeleton. Every participant signed an informed consent before the experiment. The experimental protocol received approval from the Chinese Ethics Committee of Registering Clinical Trials (ChiECRT20200319). 

\begin{figure}[!t]
	\centering
	{\includegraphics[width=\columnwidth]{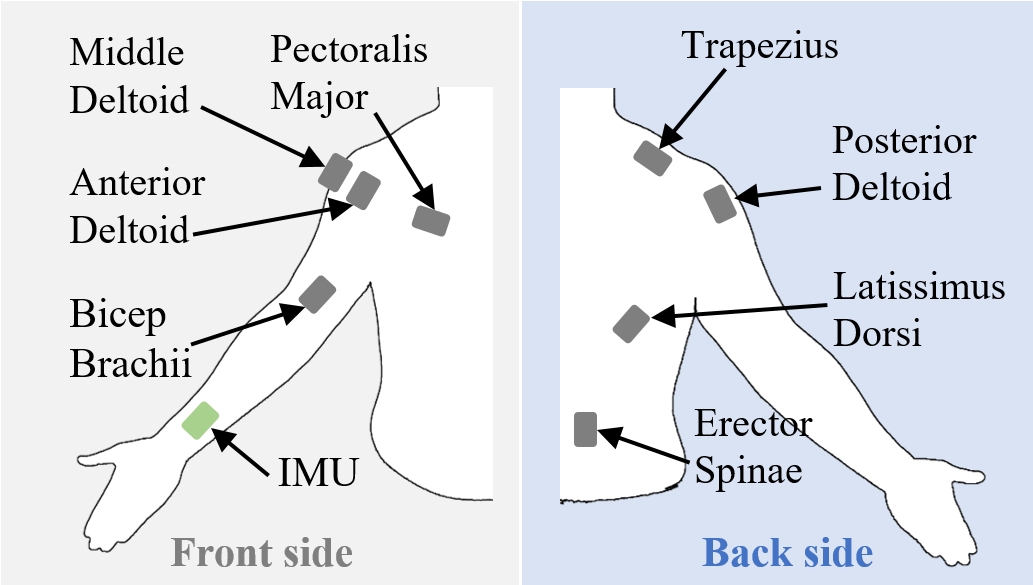}\label{fig2}}
	\captionsetup{justification=raggedright, singlelinecheck=false}
	\caption{Location of the EMG sensors and the IMU.}
\end{figure}

The participants performed the screwing task on an adjustable-height aluminum frame in two conditions (i.e., Normal and Intervention). The height of the task was defined as the height of the hand when the shoulder is raised 90°. Before the experiment, the operators ensured the exoskeleton was properly fitted to the participants and guided them to familiarize themselves with both the exoskeleton and the task. 30 minutes was provided to rest between conditions. The order of the conditions was randomized across participants to avoid order effects. Surface EMG electrodes (1111Hz, Delsys Inc., Natick, MA, USA) were unilaterally placed on right-side muscles according to SENIAM guidelines \cite{hermens2000development}: anterior deltoid (AD), middle deltoid (MD), posterior deltoid (PD), bicep brachii (BB), pectoralis major (PM), trapezius (TR), latissimus dorsi (LD), erector spinae (ES), as shown in Fig. 2(b). To normalize EMG amplitude values, the EMG signals were normalized for each participant and each task using the maximal amplitude of each muscle during Normal conditions, following the paradigm in \cite{tian2024novel}. Additionally, an IMU (148Hz, Delsys Inc., Natick, MA, USA) was placed on the right wrist to measure the acceleration (Fig. 2). 

\subsection{Data Processing}

We processed the acceleration signals to determine the task execution phases of participants and calculated the root mean square (RMS) of each muscle among participants to analyze the temporal distribution of muscle activities in both conditions.

\textbf{\textit{1) Acceleration:}} The acceleration signals were resampled to 1111Hz and variances were calculated through a non-overlapping 30ms time sliding window. We used a variance threshold of mean + 2std of the windowed data to determine the task execution phase \cite{xiaomin2020}. Specifically, the interval during which the acceleration is less than the threshold and the duration is greater than 5s was considered as the task execution phase.

\textbf{\textit{2) RMS:}} Raw EMG data were set to zero-mean, band-pass filtered (4th order low-pass Butterworth, 20-350Hz), full-wave rectified, low pass filtered (6 Hz), and normalized by the maximal amplitude of each muscle during Normal condition. Subsequently, the EMG envelope was obtained. The RMS of each muscle was obtained from the envelope via non-overlapping 30ms time sliding windows during the task execution phase. 

\subsection{Muscle Synergy}
To explore the effect of the exoskeleton on motor patterns, we utilized muscle synergy analysis to investigate the spatial distribution of muscle activities. Here, we formulated descriptions of NMF and EMG topographic maps for the screwing task in two conditions (i.e., Normal and Intervention) based on the monitored eight-channel muscle signals during the task execution phase.

\subsubsection{NMF}
In this study, we performed NMF separately for each subject and each condition. The EMG envelopes of each participant ($E$) can be formulated as \cite{rabbi2020non}:

\begin{equation}
	E=W \times H,
	\label{eq1}
\end{equation}
where $E$ is a matrix with dimensions $m \times n$, $m$ = 8 is the number of muscles, $n$ represents the number of data points; $W$ represents muscle synergy matrix with dimensions $m \times s$, $s$ is the number of muscle synergies; $H$ is an activation coefficient matrix with dimensions $s \times n$. Each element of $W$ denotes the weight for a muscle in a synergy. Each row of $H$ denotes the changes of the synergy activation over time. Normalized muscle synergy matrix $W_{norm}$ were normalized by summation across all the elements of $W$, each element of which represents the contribution of different muscles in each synergy. Normalized activation coefficient matrix $H_{norm}$ was normalized by maximum across all the elements of $H$, each row of which reflects the intensity of synergy activation, as described in (2) and (3).

\begin{equation} 
	W_{\text{norm}}(i,j) = \frac{W(i,j)}{\sum_{k=1}^{m} W(i,k)},
	\label{eq2} 
\end{equation}

\begin{equation} 
	H_{\text{norm}}(j,l) = \frac{H(j,l)}{max(H(j,:)},
	\label{eq3} 
\end{equation}
where $i \in {1,2,\ldots,m}$, $j \in {1,2,\ldots,s}$, and $l \in {1,2,\ldots,n}$. We varied the number of synergies from 2 to 8 and selected the minimum number of synergies when the variance accounted for (VAF) of all muscles is greater than 90\% and that of an individual muscle is greater than 75\% \cite{chvatal2013common} \cite{steele2015muscle}. 

To distinguish the differences in muscle synergies under the two conditions, we observed changes in muscle contribution (i.e., $W_{norm}$) to identify alterations in motion patterns. In addition, we calculated the averaged recruitment level ($Recr$) and the percent of activation duration ($Ad$) of each synergy to quantitatively compare the normalized activation coefficient matrix ($H_{norm}$). $Recr$ was obtained by averaging the $H_{norm}$ during the task execution phase. $At$ represents the ratio of the duration to the total duration of the task execution phase, where the amplitude of $H_{norm}$  greater than 0.5 is considered obvious activation \cite{flanders2002choosing}, as described in (4) and (5).

\begin{equation} 
	Recr(j) = \frac{\sum_{l=1}^{n} H_{\text{norm}}(j,l)}{n},
	\label{eq4} 
\end{equation}

\begin{equation} 
	At = \frac{\sum_{l=1}^{n} XOR(H_{\text{norm}}(j,l) > 0.5)}{n},
	\label{eq5} 
\end{equation}
where $j \in {1,2,\ldots,s}$, $l \in {1,2,\ldots,n}$ and $XOR(H_{\text{norm}}(j,l) > 0.5)$ denotes the $OR$ operation across the rows of $H_{norm}$ for all the elements larger than 0.5. That is, we first set each element of $H_{norm}$ as 1 if it is larger than 0.5 or as 0 otherwise. Then, we performed $OR$ operation across the rows and got one $1*n$ vector whose elements are 1 or 0. And we summed across the elements of the vector and then divided the summation by $n$  to get $At$.

To better understand whether the assistance of the shoulder exoskeleton induces the merging and fraction of muscle synergies, we performed non-negative regression \cite{cheung2020plasticity} to regress the columns of $W_{norm}$ during Normal conditions against each column of $W_{norm}$ during Intervention conditions. The linear coefficients of the regression denote the merging of muscle synergies. We calculated the goodness of fit ($R$), synergy similarity ($S_s$) and synergy similarity variance ($\sigma_{S_s}$). Moreover, we calculated the synergy sparseness ($S_{sp}$) through the muscle synergy matrix ($W$) in both conditions during the task execution phase. The formula can be described in (6) to (9).

\begin{equation} 
	R = \frac{1}{n_1}\sum_{i=1}^{n_1}(W_1[i]\cdot(\sum_{j=1}^{m}c_j[i]\cdot W_2[j])),
	\label{eq6} 
\end{equation}

\begin{equation} 
	S_s = \frac{1}{n_1n_2}\sum_{i=1}^{n_1}\sum_{j=1}^{n_2}c_j[i],
	\label{eq7} 
\end{equation}

\begin{equation} 
	\sigma_{S_s} = \frac{1}{n_1}\sum_{i=1}^{n_1}std(c[i,:]),
	\label{eq8} 
\end{equation}

\begin{equation} 
	S_{sp} = \frac{\sqrt{m}-\frac{\sum_{i=1}^{n}|W_i|}{\sqrt{\sum_{i=1}^{n}{W_i}^2}}}{\sqrt{m}-1},
	\label{eq9} 
\end{equation}
where $n_1$ and $n_2$ are the numbers of muscle synergies in the Normal and Intervention conditions during the task execution phase, respectively; $m$ is the number of muscles; $W_1\in \mathbb{R}^{n_1 \times m})$ and $W_2\in \mathbb{R}^{n_2 \times m})$ are the muscle synergy matrix in the Normal and Intervention conditions during the task execution phase, respectively; $c_j[i]$ represents the fitting coefficient between the $i-th$ and $j-th$ muscle synergies. $R$ measures the fit between $W_1$ and $W_2$, a higher $R$ value indicates greater similarity in muscle synergies between the two conditions. $S_s$ represents the overall similarity between $W_1$ and $W_2$ during the fitting process, a higher $S_s$ value indicates the exoskeleton has a smaller impact on the muscle synergy. $\sigma_{S_s}$ measures the variance across different muscle synergies, a higher $\sigma_{S_s}$ value indicates greater changes in the similarity between the different muscle synergies. $S_{sp}$ denotes the degree of concentration of active muscles within each muscle synergy, a higher $S_{sp}$ value indicates that fewer muscles are relied upon or that the activation intensity is lower within the synergy.

\subsubsection{EMG Topographic Map}
After 8 channels of EMG signals were preprocessed, the RMS in the two conditions across all participants were extracted during the task execution phase. Then we used linear cubic spline interpolation to standardize all RMS across all participants and all conditions during the task execution phase to the same data length, the minimum value among all RMS data lengths. For the convenience of visual comparison of differences between the two conditions (i.e., Normal and Intervention), we normalized all RMS features by selecting the maximum value among them. Then, the RMS features in each condition are averaged across all individuals. Finally, we stacked the RMS features over time and over EMG channels to construct the EMG topological maps separately for Normal and Intervention conditions, whose vertical axis denotes the channels of EMG and horizontal axis denotes time. To quantify the temporal and spatial distribution of muscle activity, four different indices were calculated from the EMG topological map: mean value ($Mean$), the coordinates of the center of gravity in the horizontal ($CoG_x$) and vertical ($CoG_y$) directions, and the entropy ($Entropy$) of EMG topological map. They can be computed as follows \cite{jiang2024effect}:

\begin{equation} 
	Mean = \frac{\sum{t(i)}}{N},
	\label{eq10} 
\end{equation}

\begin{equation} 
	CoG_x = \frac{\sum{t(i)x(i)}}{\sum{t(i)}},
	\label{eq11} 
\end{equation}

\begin{equation} 
	CoG_y = \frac{\sum{(t(i)y(i)})}{\sum{t(i)}},
	\label{eq12} 
\end{equation}

\begin{equation} 
	Entropy = -\sum{p(i) log_2p(i)},
	\label{eq13} 
\end{equation}
where $N$ is the total number of data points in the EMG topological map, $t(i)$ represents the value of the $i-th$ data point, $x(i)$ and $y(i)$ represents the horizontal and vertical coordinate values of the $i-th$ data point, respectively, $p(i)$ is the value of the $i-th$ data point normalized by the summation of the EMG topological map. The $mean$ value is a measure of the average level of exertion by all monitored muscles during the task execution phase. The $CoG_x$ represents the stage of the task execution where muscle activities are primarily concentrated, i.e., the overall activation pattern of the monitored muscles. A higher $CoG_x$ value suggests that the muscle activities are concentrated more towards the later stages of the task execution. The $CoG_y$ indicates which muscle groups are more active during the task execution phase. A higher $CoG_y$ value suggests that the muscle activities are concentrated in the muscle groups at the bottom of the chart. $Entropy$ is a measure of uniformity, indicating the degree of homogeneity of muscle activation. A higher $entropy$ value corresponds to a more uniform distribution of RMS values in the topographic map \cite{li2023sample}.

\subsection{Statistical Analysis}
Statistical analyses were used to assess the effectiveness of the exoskeleton in muscle activation and muscle synergy. The Shapiro-Wilk test was employed to check data normality. If the normal distribution was not satisfied, the Wilcoxon signed-rank test was used to perform statistical analysis, otherwise, the t-test was utilized. Statistical significance was concluded when p \textless 0.05.

\section{Results}

\subsection{Muscle Synergy in NMF}

\begin{figure}[!t]
	\centering
	{\includegraphics[width=\columnwidth]{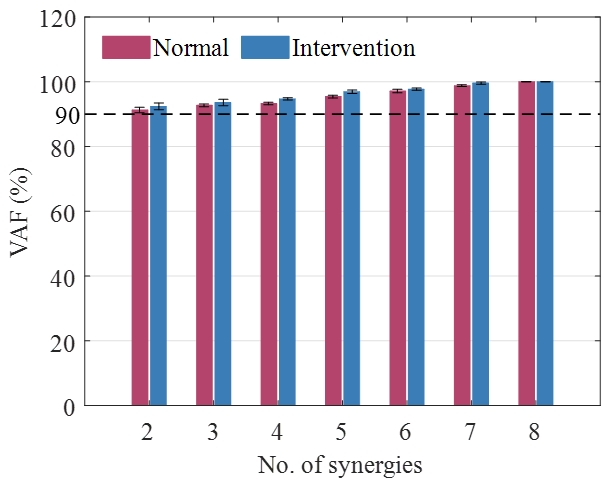}\label{fig3}}
	\caption{The variability accounted for (VAF) based on different numbers of synergies for the screwing task.}
\end{figure}

\subsubsection{Number of Muscle Synergies Across Participants}
Fig.3 shows the VAF for different numbers of synergies across all participants in the two conditions for the screwing task. VAF is commonly used in muscle synergy analysis to quantify the performance of the NMF-based muscle synergy factorization. As muscle synergies increase, VAF increases, indicating better reconstruction capability \cite{afzal2022evaluation}. In both conditions, no significant difference was found in VAF, and the optimal number of muscle synergies in both conditions was determined to be 2 based on the selection criterion. This indicates that wearing the exoskeleton may not alter the number of muscle synergies, i.e., the complexity of muscle cooperation.

\begin{figure*}[!t]
	\centering
	{\includegraphics[width=18cm]{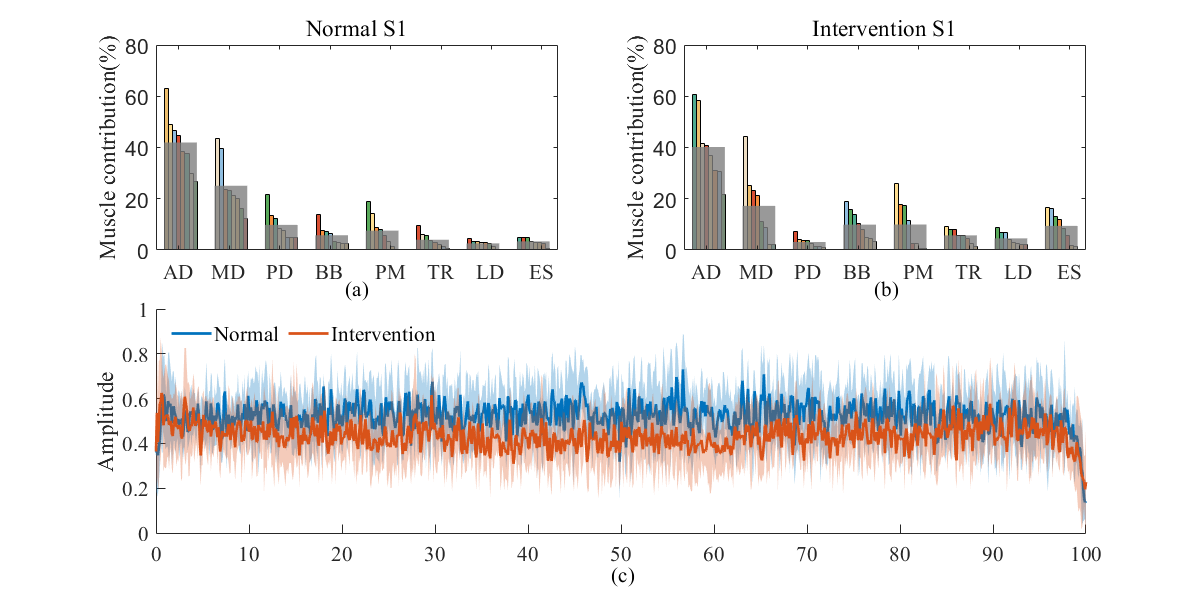}\label{fig4a}}
	
	{\includegraphics[width=18cm]{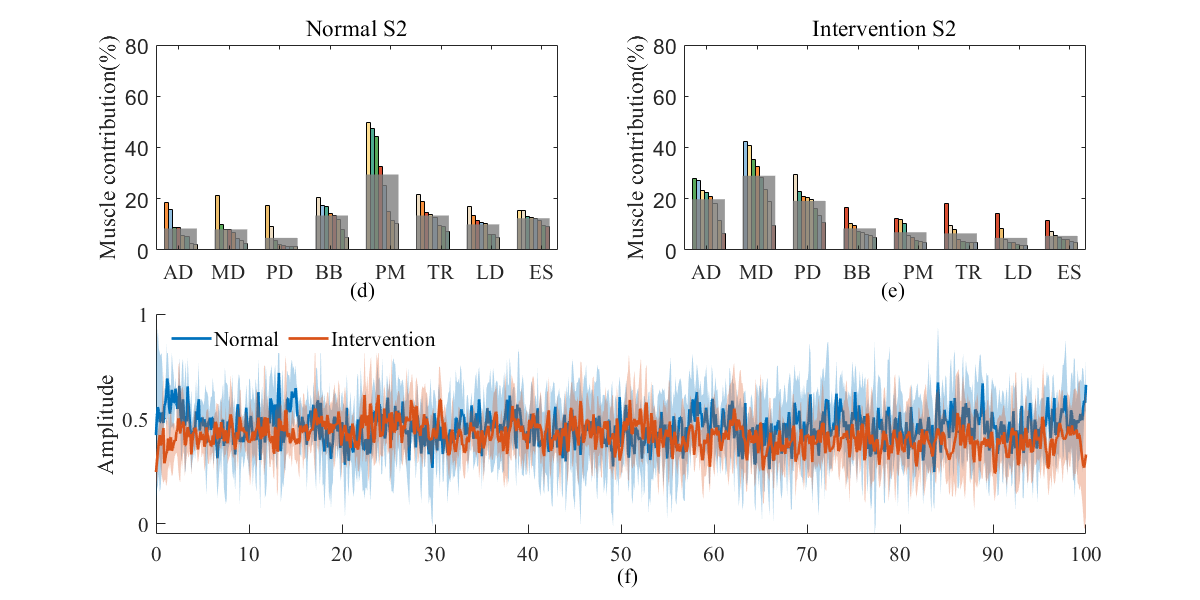}\label{fig4b}}
	
	\caption{Muscle synergies in the Normal and Intervention conditions for the screwing task. (a) The first synergy in the Normal condition. (b) The first synergy in the Intervention condition. (c) The mean of the normalized activation profiles ($H_{norm}$) across all participants in the first synergy during the two conditions. (d) The second synergy in the Normal condition. (e) The second synergy in the Intervention condition. (f) The mean of the normalized activation profiles ($W_{norm}$) across all participants in the second synergy during the two conditions. The bars of the same color represent the same participant, while the gray bars represent the mean value of the participants. The light blue and orange shaded regions represent the standard error of $W_{norm}$ during the Normal and Intervention conditions, respectively.}
\end{figure*}

\subsubsection{Muscle Synergy Identification}
The NMF algorithm provides the muscle synergy and activation coefficient but does not specify their corresponding order \cite{afzal2022evaluation}. Consequently, we performed the k-means clustering algorithm on muscle synergies (i.e. each column of $W$) from all the subjects and clustered then into two groups, in order to sort the two synergies for each subject \cite{ghislieri2020muscle}. After post-hoc check, we found that the two synergies of all the subject were divided into two separate categories. The synergy with the highest weight of AD and MD muscle (shoulder agonist muscles) was named as the first synergy, and the other synergy was assigned to the second one. We defined the two muscle synergies under the two conditions as Normal\underline{~}S1, Normal\underline{~}S2, Intervention\underline{~}S1, and Intervention\underline{~}S2. Specifically, for Normal\underline{~}S1 and Intervention\underline{~}S1, the muscle synergy is similar across participants. The synergy except for S1 is S2. After visually checking the primarily activated muscle in S2 of each subject and each task, we found that for Normal\underline{~}S2 and Intervention\underline{~}S2, 6 participants in Normal condition and 7 participants in Intervention condition present similar muscle synergy across participants. A representative set of muscle synergies across all participants and the corresponding activation profiles are demonstrated in Fig. 4. Normal\underline{~}S1 and Intervention\underline{~}S1 primarily involve AD and MD. The main contributions to Normal\underline{~}S2 and Intervention\underline{~}S2 come from PM and MD, respectively. There was a strong correlation between Normal\underline{~}S1 and Intervention\underline{~}S1 ($r$ = 0.94), but Normal\underline{~}S2 and Intervention\underline{~}S2 showed opposite correlation ($r$ = -0.45).

\subsubsection{Activation Profiles}
We compared the mean of the normalized activation profiles across all participants in each muscle synergy between the two conditions, as shown in Fig. 4. We averaged each activation profile (i.e. each row of the matrix $H_{norm}$) across time and compared the averages of all the subjects across the two conditions. For the first synergy, significant differences were observed between the Normal\underline{~}S1 and Intervention\underline{~}S1 ($p$ = 0.0371). As shown in Fig. 4(c), the amplitude of the activation profile during the mid-phase (30\% - 70\%) in the Intervention condition was lower than that in the Normal condition, while the amplitude of the activation profile in the late phase (90\%-100\%) did not exhibit an increasing trend. In terms of the second synergy, there was not a significant difference between the Normal\underline{~}S2 and Intervention\underline{~}S2 ($p$ = 0.958). As shown in Fig. 4(f), the amplitude of the activation profile in the Intervention condition did not show a substantial decrease compared to the Normal condition.

\subsubsection{Recr and Ad}
The averaged recruitment level ($Recr$) in Intervention\underline{~}S1 was lower than that in Normal\underline{~}S1, which exhibited statistically significant (Normal vs. Intervention: 50±11\% vs. 33±9\%, $p$=0.03); that in InterventionS2 and NormalS2 presented the same trend, but no significance (Normal vs. Intervention: 40±16\% vs. 32±11\%, $p$=0.13), as shown in Fig. 5(a). Compared to the Normal condition, the percent of activation duration ($Ad$) in the Intervention condition showed a significant reduction in the first synergy (Normal vs. Intervention: 65±10\% vs. 40±24\%, $p$=0.04), while there was no significant difference in the second synergy (Normal vs. Intervention: 28±13\% vs. 24±7\%, $p$=0.47), as shown in Fig. 5(b). 

\begin{figure}[!t]
	{\includegraphics[width=\columnwidth]{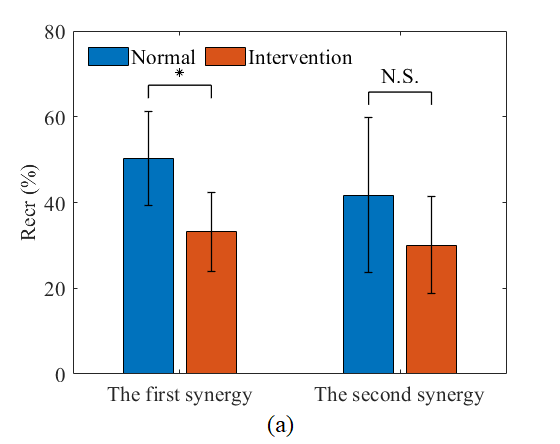}\label{fig5a}}
	{\includegraphics[width=\columnwidth]{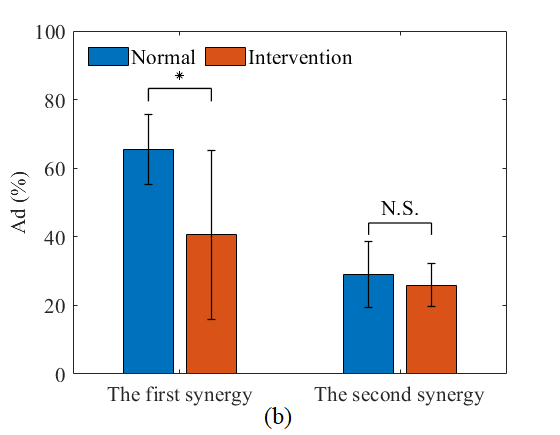}\label{fig5b}}
	\caption{The averaged recruitment level ($Recr$) and the percent of activation duration ($Ad$) of each synergy in the Normal and Intervention conditions for the screwing task. (a) $Recr$. (b) $Ad$. * denotes a significant difference ($p$\textless0.05). N.S. stands for “No Significant”.}
\end{figure}

\subsubsection{Similarity and Sparseness}
For the metrics obtained by the non-negative regression (i.e. $R$, $S_s$ and $\sigma_{S_s}$), we constructed the null distribution by regress the columns of $W_{norm}$ during Normal conditions against the columns of $W_{norm}$ during Intervention condition of other subjects. By comparing the metrics with the metrics obtained by the null distribution, we can get $p$ values and evaluate whether the metrics are beyond random. The averaged goodness of fit ($R$) is 0.1896, where 6 out of 8 subjects showed a goodness of fit significantly larger than random. As for the synergy similarity ($S_s$), the average is 0.4607 and 4 out of 8 subjects presented a significant effect compared with the null distribution. In terms of the synergy similarity variance ($\sigma_{S_s}$), the average is 0.4371 and 2 out of 8 subjects presented significantly larger synergy similarity variance over random. The average synergy sparseness $S_{sp}$ is 0.4023 for Normal conditon and 0.3871 for Intervention condition. It did not present a significant difference between the two conditions ($p$ = 0.7341).

\subsection{Muscle Synergy in EMG Topographic Map}

The EMG topographic maps in both conditions depicted from another perspective of how the muscles coordinate with each other across time (Fig. 6). In both conditions, the high activity region was displayed in AD and MD. Moreover, the ES also showed high activity in the Normal condition. According to the color bar coordinates, the muscle activation of all monitored muscles generally decreased when wearing the exoskeleton. In quantitative analysis results (Fig. 7), the results of the mean value in the Intervention condition revealed a statistically significant decrease concerning the Normal condition (Normal vs. Intervention: 42±5\% vs. 27±4\%, $p$\textless 0.001). The coordinates of the center of gravity in the horizontal direction ($CoG_x$) and vertical ($CoG_y$) directions, no statistically significant changes were found between conditions (Fig. 7(b)-(c)). There was a significant difference in entropy between conditions (Normal vs. Intervention: 8.76±0.75 vs. 11.04±1.00, $p$=0.0012), as shown in Fig. 7(d).

\begin{figure*}[!t]
	{\includegraphics[width=\columnwidth]{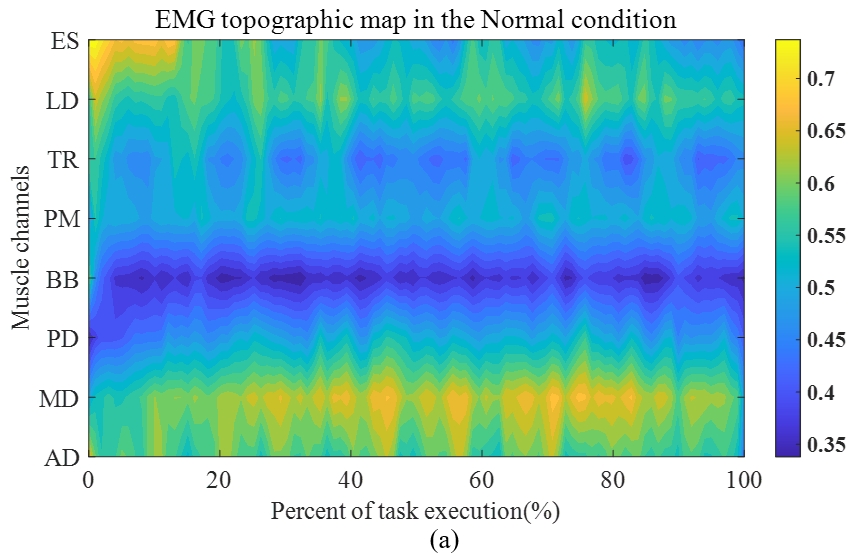}\label{fig6a}}
	{\includegraphics[width=\columnwidth]{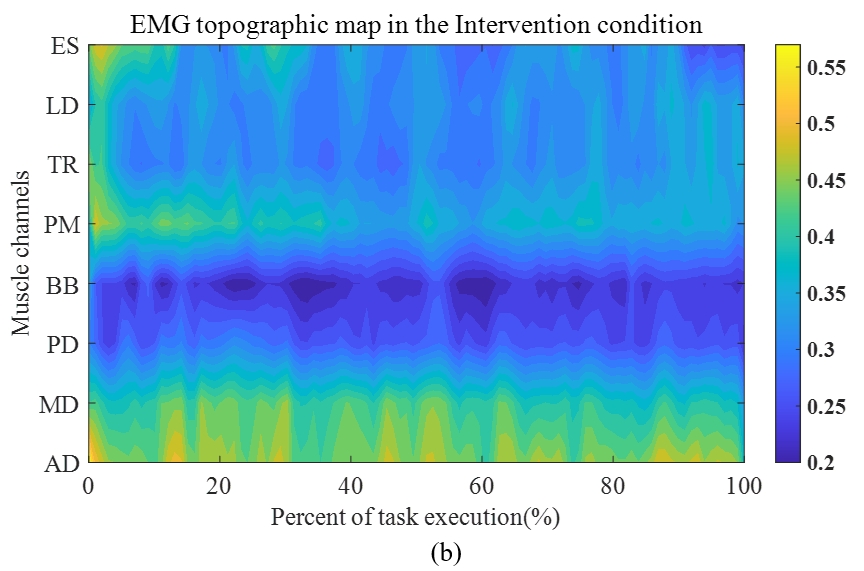}\label{fig6b}}
	\caption{The EMG topographic maps in both conditions for the screwing task. (a) The Normal condition. (b) The Intervention condition.}
\end{figure*}

\begin{figure*}[!t]
	\centering
	{\includegraphics[width=15cm]{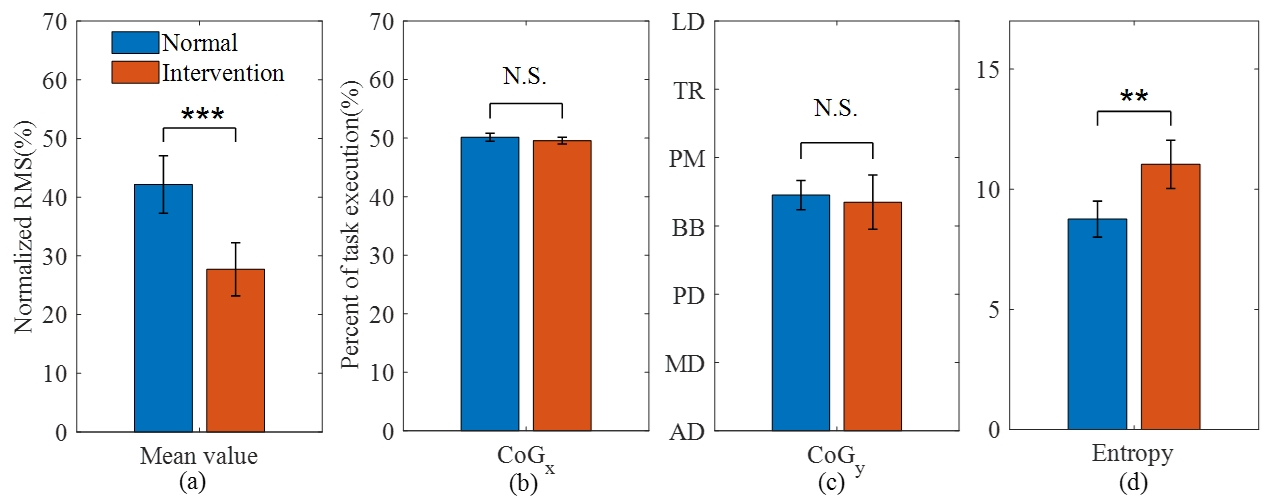}\label{fig7}}
	\caption{The evaluation indicators calculated from the EMG topographic maps in both conditions for the screwing task. (a) The $mean$ value. (b) The coordinate of the center of gravity in the horizontal ($CoG_x$) direction. (c) The coordinate of the center of gravity in the vertical ($CoG_y$) direction. (d) The $entropy$. *** and ** denotes significant differences ($p$\textless0.001 and $p$\textless0.01, respectively). N.S. stands for “No Significant”.}
\end{figure*}

\section{Discussion }

\subsection{Summary}
We investigated the impact of the exoskeleton on muscle synergies. We found that the exoskeleton could change the muscle coordination of participants to make them more efficient (i.e., lower recruitment levels and less activation time), without increasing its complexity (i.e., the total number of muscle synergies remains unchanged). Therefore, it has the potential to alleviate fatigue and reduce the incidence of shoulder WMSDs. However, it also could induce new muscle synergies while hiding or weakening existing non-primary synergies, without altering the primary ones.

\subsection{Muscle Synergy in NMF}

\subsubsection{Number of Muscle Synergies Across Participants}
 We observed no significant difference in the number of muscle synergies between conditions. This indicates that wearing the exoskeleton would not increase the complexity of the overhead work, that is, it would not bring additional burden to the nervous system \cite{afzal2022evaluation}. This result aligned with our hypothesis. The variation in the number of muscle synergies typically occurs under different neurological conditions, such as cerebral palsy, due to alterations caused by neurological disorders \cite{steele2015muscle}. In this study, the assistance provided by the exoskeleton was the dominant variable faced by the participants; therefore, the lack of variation in the number of muscle synergies is acceptable.

\subsubsection{Muscle Synergies}

For the first synergies in both conditions, they may correspond to the screwing action due to the following reasons. First, the primary contributing muscles of the first synergy were AD and MD, which are primarily responsible for shoulder flexion. Second, the averaged recruitment level and percent of activation duration of the first synergy is larger than that of the second synergy (Fig 5). Third, the screwing task was continuous, that is, lifting the arm and tightening all screws before lowering the arm, which reflected the fact that screwing action lasted longer in the task than the transitions between two screwing actions. For the second synergies in both conditions, the muscles contributing the most were PM and MD (Fig. 4(d) - (e)), respectively, which are responsible for shoulder stabilization \cite{grazi2020design}. Thus, the second synergy may be responsible for shoulder stability. 

Regarding the first synergies in both conditions, they can be considered the same synergy due to the strong correlation ($r$=0.94) between the first synergies in both conditions. This suggests the exoskeleton assistance does not alter the existing primary muscle synergy (i.e., the first synergy). All participants (8 out of 8) exhibited the first synergy across both conditions. The contributions of AD, MD, and PD in the Intervention condition were lower than those in the Normal condition. The changes could be due to the assistance provided by the exoskeleton, reducing the burden on the shoulder agonist muscles, which is consistent with previous studies \cite{kim2022passive} \cite{pacifico2022exoskeletons}.

Regarding the second synergies in both conditions, we observed variations in the number of participants involved in the second synergy, with 6 out of 8 participants displaying it under the Normal condition, compared to 7 out of 8 under the Intervention condition. This observation reflects individual differences, indicating that each participant’s familiarity with the task and adaptation to the exoskeleton may differ. Moreover, the second synergies in the two conditions showed weak correlation ($r$=-0.45) and involved different primary muscles (PM and MD in the Normal and Intervention conditions, respectively), indicating they can be considered two different synergies. This phenomenon indicates that the exoskeleton may induce a new synergy to emerge, aligning with the concept of muscle synergy plasticity \cite{cheung2020plasticity} \cite{clark2010merging}. Similar results can be found in exoskeleton-assisted walking \cite{afzal2022evaluation}. 

For the second synergy in the Normal condition, the proportion of the PM was much higher than other muscles, which may be related to human behavioral patterns, suggesting that PM may have a compensatory effect during the task. For the second synergy in the Intervention condition, the primary muscle had changed from PM to MD. The reason for this transition could be attributed to the assistance provided by the exoskeleton, which reduced the physical effort on shoulder muscles, consequently decreasing the reliance on PM. In addition, the increase in the contribution of MD is likely due to the potential pressure exerted on it by the torque generator placed on the side of the arm, thereby increasing its muscle activation to cope with the inadaptability induced by external stimuli. This suggests that we need to pay particular attention to the user-friendly and comfortable interface of the exoskeleton during the design phase.

\subsubsection{Activation Profiles}

In terms of the first synergy, we found a significant decrease in the amplitude of the activation profile in the Intervention condition. The observation may be attributed to the assistance provided by the exoskeleton, which resulted in the lower amplitude of the activation profiles in the Intervention condition. In addition, compared to the Normal condition, there was no increasing trend in the activation profile during the late phase (90\%-100\%), which includes the process of lowering the arm. This could be explained by the small assistance at low elevation angles provided by the exoskeleton, providing evidence that the exoskeleton does not hinder human movements \cite{tian2024novel}. In terms of the second synergy, we found no significant difference in the activation profiles between the two conditions. This phenomenon is consistent with the results of the average recruitment level ($Recr$) and the percentage of activation time ($Ad$) in the second synergy, as shown in Fig. 5.

\subsubsection{Recr and Ad}
Regarding the first synergy, both the average recruitment level ($Recr$) and the percentage of activation time ($Ad$) in the Intervention condition were significantly reduced compared to the Normal condition. They represent the manifestation of muscle synergy control strategy, providing information on the involvement level and duration of muscles \cite{ghislieri2020muscle}. The results indicate that wearing the exoskeleton could reduce the burden (i.e., activation intensity and duration) on muscles and offer a potential delay in muscle fatigue, especially the shoulder agonist muscles (e.g., AD and MD). We attribute the difference to the compensation of shoulder gravity provided by the exoskeleton assistance. Similar results have been observed in existing studies on muscle activation \cite{de2023passive} \cite{van2022exo4work}. Regarding the second synergy, there were no significant differences in the $Recr$ and $Ad$ between the two conditions. The results are a quantitative response to the amplitude of activation profiles in Fig. 4(f).

\subsubsection{Similarity and Sparseness}
The metrics of similarity and sparsity presented some insights into how the different muscle synergies emerge under the assistance of the exoskeleton. The generally above-random goodness of fit ($R$) indicated that the muscles synergies of Intervention condition is caused by decomposing and merging the original muscle synergies, i.e. the muscle synergies of Normal condition. The synergy similarity ($S_s$) suggests a certain degree of similarity in muscle synergies between the two conditions, likely due to the high similarity of the first synergy in both conditions. Although 4 out of 8 subjects present significantly difference on $S_s$, 2 of the 4 subjects present smaller values and the rest 2 subjects present larger values compared with random. This suggests the individual difference among subjects for adapting to the assistance of exoskeleton. Only 2 out of 8 subjects present a significantly larger synergy similarity variance ($\sigma_{S_s}$) over random. This, combining with the phenomena of no significant difference on synergy sparseness ($S_{sp}$) across two conditions, suggests that although the assistance of exoskeleton induces a change of muscle synergies, it does not optimize the muscle coordination, i.e. the neural drive strategies of multiple muscles \cite{cheung2020plasticity}.

\subsection{Muscle Synergy in EMG Topographic Map}

Through visual inspection of the EMG topographic maps, it is evident that the overall amplitude in the Intervention condition was generally lower compared to the Normal condition. This suggests a potential delay in muscle fatigue, which aligns with our NMF-based analysis of muscle synergies. During the task execution phase (i.e., along the horizontal axis), there were no significant changes from blue to yellow in the regions corresponding to each muscle in both conditions, indicating that the participants did not reach a state of fatigue since fatigue would induce increasing RMS of corresponded muscles. 

The mean value in the Intervention condition exhibited significant changes with respect to the Normal condition. It can be considered as another representation of muscle activation \cite{jiang2024effect}. Such a positive impact on mean values may be due to the fact that the exoskeleton provided effective assistance. This is consistent with previous studies about muscle activation \cite{kim2022passive} \cite{maurice2019objective}. We observed that entropy showed a significant difference in the Intervention condition compared to the Normal condition. This may be supported by the observed reduction in amplitude of the EMG topographic maps, as well as in the average recruitment level ($Recr$) and the percentage of activation time ($Ad$) of the first synergy identified using NMF between the two conditions. For muscle synergies in NMF and EMG topographic maps, we can observe that they complement each other and mutually confirm each other's results, thus demonstrating the validity and rationality of our analytical approach. The entropy can be used to measure the uniformity of values, suggesting the degree of homogeneity in muscle activation \cite{jiang2024effect}\cite{farina2008change}. A higher value of the entropy represents a more uniform distribution \cite{jiang2024effect} \cite{farina2008change}. This finding suggests that the changing pattern of the heterogeneity of the distribution of monitored muscles would be influenced by the exoskeleton. In terms of the $CoG_x$ and $CoG_y$, there were no significant differences between conditions. This suggests that wearing the exoskeleton did not alter timing of muscle exertion, nor did it change the main active muscle groups, which are consistent with the previous research \cite{jiang2024effect}. However, we observed that $CoG_y$ was lower in the Intervention condition, indicating reduced activities in the AD, MD, PD, and BB muscle groups. It may suggest that the assistance provided by the exoskeleton reduced the load on the related muscle groups during the task. 

\subsection{Comparison with Previous Studies on Muscle Synergy}

We compared our work with studies utilizing NMF and EMG topographical maps to analyze the impact of the exoskeleton on muscle synergy, as well as studies on the assessment of the shoulder exoskeleton, aiming to deepen the understanding of muscle synergy during exoskeleton-assisted overhead work.

Firstly, we compared our muscle synergies analysis using NMF with the study by Afzal and colleagues \cite{afzal2022evaluation}. They evaluated the muscle synergy during exoskeleton-assisted walking in persons with multiple sclerosis. They found that exoskeleton assistance did not alter the existing synergies, but could induce a new synergy to emerge \cite{afzal2022evaluation}. But we found that the exoskeleton did not alter the existing primary synergy, but would alter non primary synergy. This may be because, with the assistance of the exoskeleton, the original synergy was insufficient to fully interpret the original EMG signals, leading to the emergence of a new synergy \cite{cheung2020plasticity}. This also suggests that the central nervous system employs flexible control over muscle synergy. Additionally, compared to the walking of persons with multiple sclerosis, the overhead work in healthy persons involves fewer muscle groups and simpler synergy patterns. This could also be one of the reasons for the observed differences of the results between our and Afzal’s study \cite{afzal2022evaluation}. 

Secondly, we compared our muscle synergies analysis using EMG topographic maps with the study conducted by Jiang and colleagues \cite{jiang2024effect}. They analyzed the synergistic effects of lower back muscles when wearing a lumbar exoskeleton and found that the entropy did not show a significant difference among conditions in the lifting task. However, we observed a higher entropy value in the Intervention Condition. The contrasting results could be attributed to that they only monitored the lower back muscle group, excluding other muscles involved in the task, while we monitored a greater number of muscles involved in the task. Furthermore, in the holding task, they found the entropy showed a difference between trials, and thus they concluded that the active lumbar exoskeletons can reduce the load on low back muscles in the static holding task rather than in the dynamic lifting task \cite{jiang2024effect}[38]. We investigated the muscle synergy effect on the screwing task, which was a quasi-static task, hence the results may be more consistent with those of a holding task. 

Finally, compared with previous studies about assessment of the shoulder exoskeleton \cite{kim2022passive} \cite{de2022occupational}. To our knowledge, we are the first to investigate the impact of the exoskeleton on muscle synergy during overhead work, although this is just a case study. Our muscle synergy analysis results provide an interpretation for previous research findings, aiding in a deeper understanding of human motor mechanisms and the assistive mechanisms of exoskeletons. We found that the exoskeleton does not alter the number of synergies and existing primary synergies, but it may induce the emergence of new synergies. We should pay much attention to the friendly interaction between exoskeletons and humans to prevent undesired synergies. Additionally, we should also focus on individual differences in movement habits (i.e., different muscle synergies) so that exoskeletons can adapt to a broader range of users.

\subsection{Limitations}
The study presented in this paper has some limitations. The participants were selected as healthy males, all right-handed, and the sample size was relatively small. Despite the positive outcomes of the muscle synergy analysis, other factors such as including females, left-handed individuals, a wider age range, and experienced workers should be considered in the design of future experiments. Another limitation of this study is the limited variety and complexity of overhead tasks, resulting in a small number of synergies using NMF. As this study is a case study, only the common screwing task among overhead tasks was selected. In future research, a wider variety and more complexity of overhead tasks should be considered to investigate the impact of exoskeletons on muscle synergy. Additionally, the effect of exoskeletons on muscle synergy during overhead tasks under the fatigue state of users is also worthy of investigation. Finally, the present study demonstrated a limited correlation between muscle synergies and neurophysiology.

\section{Conclusion}
This study investigated the impacts on muscle synergies of the exoskeleton during overhead work. There are three major findings about muscle synergies in this study. Firstly, during the overhead work, the exoskeleton may not alter the number of synergies and existing primary synergies, but it may induce new ones. Secondly, muscle activation (i.e., activation profiles) significantly decreased during the Intervention condition compared to the Normal condition. Finally, wearing the exoskeleton did not alter the primary portion of the task where muscle activities were concentrated (i.e., the overall activation pattern of the monitored muscles), nor did it change the main active muscle groups. Additionally, wearing the exoskeleton could make the distribution of muscle activation more uniform during the overhead work. We emphasize the practicality of muscle synergy analysis in extracting meaningful information on the movement patterns of the participant and muscle co-contractions while wearing the exoskeleton. The comprehensive muscle synergies results would provide additional feedback to the user and designer, assisting in evaluating the performance of the exoskeleton and making necessary changes in exoskeleton design and assistance.

\section*{Acknowledgment}

We gratefully acknowledge the participation of all individuals involved in this experiment. Their dedication and support were crucial to the completion of our research.

\bibliographystyle{IEEEtran}
\bibliography{IEEEabrv, myreferences}

\end{document}